\documentclass[12pt,epsf]{article}

\usepackage{amsmath}
\usepackage{graphicx}
\usepackage{epsfig}
\usepackage{axodraw}
\begin{document}
%%%%%%%%%%%%%%%%%%%%%%%%%%%%%%%%%%%%%%%%%%%%%%%%%%%%%%%%%%%%%%%%%%%%
%%%%%%%%%%%%%%%%%%%%%%%%%%%%%%%%%%%%%%%%%%%%%%%%%%%%%%%%%%%%%%%%%%%%
\newcommand{\be}{\begin{equation}}
\newcommand{\ee}{\end{equation}}
\newcommand{\bea}{\begin{eqnarray}}
\newcommand{\eea}{\end{eqnarray}}
\newcommand{\ba}{\begin{array}}
\newcommand{\ea}{\end{array}}
\newcommand{\dir}[1]{\not\!{#1}}
\newcommand{\ti}[1]{\tilde{#1}}
\newcommand{\ul}{\bar{u}(p'_1)(-ie\gamma^\mu)u(p_1)}
\newcommand{\ur}{\bar{u}(p'_2)(-ie\gamma^\nu)u(p_2)}
\newcommand{\phs}{e^{\frac{i}{2}(p_1\cdot \ti{p_1'}+p_2\cdot \ti{p'_2})}}
\newcommand{\intk}{\int\frac{d^4k}{(2\pi)^4}}
\newcommand{\al}[1]{\alpha_{#1}}
\newcommand{\inta}{\int_0^1 d\al{1} d\al{2} d\al{3} \delta(1-\sum\al{i})}
\newcommand{\ekq}{e^{-ik\cdot \widetilde{q}}}
\newcommand{\n}{\nonumber}
\newcommand{\abc}{\alpha_1+\alpha_2+\alpha_3}
\newcommand{\vs}{\vspace}
\newcommand{\ct}{\cite}
\def\l{\label}
%%%%%%%%%%%%%%%%%%%%%%%%%%%%%%%%%%%%%%%%%%%%%%%%%%%%%%%%%%%%%%%%%%%%
%%%%%%%%%%%%%%%%%%%%%%%%%%%%%%%%%%%%%%%%%%%%%%%%%%%%%%%%%%%%%%%%%%%%

\topmargin 0pt

\oddsidemargin -3.5mm

\headheight 0pt

\topskip 0mm \addtolength{\baselineskip}{0.20\baselineskip}
\begin{flushright}
  SOGANG-HEP 283/01 \\
  %{\tt hep-th/xxxxxxx}
\end{flushright}
\vs{5mm}
\begin{center}
  {\Large \bf Radiative Corrections in Noncommutative QED }\\
  \vs{10mm} {\large Ki Boum
    Eom~\footnote{kbeom@physics4.sogang.ac.kr}, Sung-Shig
    Kang,~\footnote{kangss@physics4.sogang.ac.kr}
    Bum-Hoon Lee,~\footnote{bhl@ccs.sogang.ac.kr}
    Chanyong Park~\footnote{cyong21@physics4.sogang.ac.kr}}
  \vs{10mm}\\
  {\it Department of Physics, Sogang University, Seoul 121-742, Korea}\\
  \vs{30mm} {\bf ABSTRACT}
\end{center}
We study the radiative corrections of the noncommutative QED
%The Feynman rules vertices are slightly modified with phase
%factors. Furthermore some QCD-like diagrams,which do not appear in
%ordinary case, should be added. These noncommutative effects are
%explicitly calculated
at the one-loop level.
%Among these loop
%diagrams, concerning with the electron-photon vertex, we confirmed
%some results on some noncommutative effects of the
A correction of the magnetic dipole moment due to the noncommutativity
are evaluated.  As in the ordinary
QED, IR divergence is shown to vanish
%problem is investigated. Like the ordinary QED, it disappears
when we combine both the tree level Bremsstrahlung diagram and
the one-loop electron vertex function. \vs{5mm}
\begin{flushleft}
%  PACS Nos: 11.10.Ef, 11.10.Lm, 11.15.-q, 11.30.-j \\
  %\today \\
\end{flushleft}

%%%%%%%%%%%%%%%%%%%%%%%%%%%%%%%%%%%%%%%%%%%%%%%%%%%%%%%%
\newpage
%\tableofcontents
%\newpage
%==============================================%
\section{Introduction}
%==============================================%

Field theory on the noncommutative space compared to the ordinary one,
has many interesting
properties. In recent
years, there have also been much interest in the noncommutative
field theories (NCFT) related to the string theory \ct{cds,dh}.
The quantum field theory on noncommutative space can arise
naturally as a decoupled limit of open string dynamics on
D-branes with the background NS-NS $B$ field. In particular, it
was shown \ct{cds,dh} that noncommutative geometry can be
successfully applied to the compactification of M(atrix) theory
\ct{BFSS,IKKT} in a certain background.  The low energy effective
theory for D-branes in the $B_{NS}$ field background is
specifically described by a gauge theory on noncommutative space
\ct{sw}.

The noncommutative scalar field theory with $\phi^4$
interaction is analyzed in \ct{seiberg,mar,araf,suss}and
shown to be renormalizable up to two loop level.
The QED on noncommutative space has also been discussed in
\ct{hya,as,ab,ja}. In NCQED, the Feynman rules for vertices are
slightly modified with phase factors. Also, non-abelian type
diagrams are added unlike the ordinary QED case\ct{suss,hya,ja}.

In this work we consider the radiative correction to the electron
scattering with other heavy particle, muon ($e^- \mu^-
\rightarrow e^- \mu^-$) in noncommutative QED. There are two
types of radiative correction to the tree level scattering
process as in QED : loop-corrections and the bremsstrahlung.
%The lowest-order tree-level Feynman diagram
%is just process with diagrams that contain no loops. But all such
%process
%receive
%high-order contributions, known as
The one-loop {\it radiative correction} to the tree-level Feynman
diagram and the {\it bremsstrahlung} will have additional diagrams
of non-abelian type.
%that do contain loops.
%In {\it bremsstrahlung}, another source of radiative corrections
%reaction, with diagram that contain non-abelian type diagram.

We calculate the soft bremsstrahlung, photon vacuum
polarization, and electron-photon interaction vertex with the
additional non-abelian type diagrams up to one loop level in
noncommutative QED. For the vertex function of electron-photon interaction
we evaluate the anomalous dipole moment\ct{ja}.  We find
that IR divergences for the electron vertex function are
cancelled by soft bremsstrahlung in the NCQED, just like in
ordinary QED.
%and it likewise show in the NCQED.

The paper is orgnized as follows. In section 2, the
Feynman rules of the noncommutative QED are summarized.
In section 3, we compute several
bremsstrahlung diagrams in NCQED and show that IR divergences of
these diagrams in NCQED with finite noncommutativity ($\theta$)
are equal to that of the ordinary QED, in soft photon limit.
In section 4, we
find photon vacuum polarization up to one loop level in NCQED and
like the ordinary QED, there is no IR divergences in that case. We
calculate electron vertex function in section 5. And then we evaluate the
noncommutative effects on the electro-magnetic dipole
moments\ct{ja}. In section 6, we show that IR divergence of the vertex function is
the same as that in ordinary QED.
And we finish this paper with some conclusion and
discussion/
%\newpage

%=====================================================%
\section{Noncommutative QED and Feynman Rules}
%=====================================================%
%=====================================================%
%\subsubsection*{Noncommutative QED}
%=====================================================%
The  action for the noncommutative QED is given by \ct{hya}
\begin{eqnarray} \label{ac}
  S[A_{\mu}, \overline{\psi},\psi]= \int d^{D}x \bigg[
  -\frac{1}{4}\ F_{\mu\nu}
  \left(x\right)\star F^{\mu\nu}\left(x\right)
+ i\overline{\psi}\left(x\right)\gamma^{\mu}\star
  D_{\mu}\psi\left(x\right)-m\overline{\psi}\left(x\right)
  \star\psi\left(x\right)\bigg]
\end{eqnarray}
where $F^{\mu\nu}$ is
\begin{eqnarray}
  F_{\mu\nu}\left(x\right)&\equiv& \partial_{\mu}A_{\nu}
  \left(x\right)-\partial_{\nu}A_{\mu}\left(x\right)
  +ig\big[A_{\mu}\left(x\right),A_{\nu}\left(x\right)\big]_{\star}.
\end{eqnarray}
and the covariant derivative is defined by:
\begin{eqnarray}
  D_{\mu}\psi\left(x\right)\equiv\partial_{\mu}\psi\left(x\right)
  +igA_{\mu}\left(x\right)\star\psi\left(x\right).
\end{eqnarray}
 The $\star$-product between two functions $\psi$ and $\phi$ is
defined by
\begin{eqnarray}
    \psi\left(x\right)\star \phi\left(x\right)\equiv
    e^{\frac{i\theta_{\mu\nu}}{2}\ \frac{\partial}{\partial\xi_{\mu}}\
    \frac{\partial}{\partial\zeta_{\nu}}
    }\psi\left(x+\xi\right)\phi\left(x+\zeta\right)\bigg|_{\xi=\zeta=0},
\end{eqnarray}
where $\theta_{\mu\nu}$ is a real constant antisymmetric
parameter reflecting the noncommutativity of the coordinates of
{\bf R}$^D$ \cite{sw}:
\begin{eqnarray}
    [x_{\mu},x_{\nu}]=i\theta_{\mu\nu}.
\end{eqnarray}
We will consider only the spatial noncommutativity and take,
without any loss of generality, $\theta$ to lie in $(1,2)$ plane.
$\theta_{12}=-\theta_{21}=\theta$ and others are $0$. The action
(\ref{ac}) is invariant under the local  gauge transformations of
the gauge fields  and matter fields.

%=====================================================%
%\subsubsection*{Feynman rules for the NCQED}
%=====================================================%
The change of the Feynman rules for NCQED  due to the presence of
the star product is only at the vertices. The propagators are the
same as those of ordinary QED.

%Therefore the propagators for the
%free fermion, gauge and the ghost fields are the same as in the
%case of usual QED.
\begin{center}\begin{picture}(300,60)(0,0)
\SetColor{Black} \ArrowLine(110,30)(50,30) \Text(80,20)[]{p}
\Text(180,30)[]{=\Large{$\quad {i\over{\not p-m+i\epsilon}}$}}
\end{picture}
\end{center}
\begin{center}
\begin{picture}(300,70)(0,0)
\SetColor{Black} \Photon(50,50)(110,50){4}{4}
\Text(40,50)[]{$\mu$}\Text(120,50)[]{$\nu$}
\LongArrow(90,40)(70,40)\Text(80,30)[]{q}\Text(180,50)[]{=\Large
{$\quad
    {g^{\mu\nu}\over{i(q^2+ie)}}$}}
\end{picture}
\begin{picture}(300,80)(0,0)
\SetColor{Black} \DashArrowLine(110,50)(50,50){2}\Text(80,40)[]{p}
\Text(180,50)[]{=\Large{$\quad
    {g^{\mu\nu}\over{i(p^2+ie)}}$}}
\end{picture}\\
{\sl Propagator in NCQED}
\end{center}

On the other hand, the Feynman rules for the vertices carry extra
phase factors coming from the noncommutative star products as
follows (with the notation $\tilde{p}_i =\theta^{ij}p_j$).
%%a commutative non-abelian gauge theory with all matter fields in
%the adjoint of the gauge group. So we can see the following
%Feynman rules.
\begin{center}
\begin{picture}(300,90)(0,0)
\SetColor{Black} \ArrowLine(30,60)(70,50)\Text(50,45)[]{$p_i$}
\ArrowLine(70,50)(110,60)\Text(90,45)[]{$p_f$}
\Photon(70,20)(70,50){2}{3}\Text(80,30)[]{$p_\gamma$}
\Text(70,10)[]{$\mu$}\Text(195,50)[]{=\quad{$ie\gamma^\mu
    e^{{i\over2}p_i\tilde p_f}$}}
\end{picture}
\begin{picture}(400,150)(0,0)
\SetColor{Black}
\Photon(120,80)(120,50){2}{3}\Text(120,90)[]{$\mu_1$}\Text(130,65)[]{$p_1$}
\ArrowLine(120,63)(120,67)
\Photon(80,40)(120,50){2}{4}\Text(100,35)[]{$p_2$}\Text(70,30)[]{$\mu_2$}
\ArrowLine(102,46)(98,44)
\Photon(120,50)(160,40){2}{4}\Text(140,35)[]{$p_3$}\Text(170,30)[]{$\mu_3$}
\ArrowLine(138,44)(142,46)
\Text(260,110)[]{=\quad{$-2e\sin\Big({1\over2}p_1\tilde
    p_2\Big)$}}
\Text(255,80)[]{\quad\quad\quad{$\times\bigg[(p_1-p_2)^{\mu_3}g^{\mu_1\mu_2}$}}
\Text(255,50)[]{\quad\quad\quad{$+(p_2-p_3)^{\mu_1}g^{\mu_2\mu_3}$}}
\Text(255,20)[]{\quad\quad\quad{$+(p_3-p_1)^{\mu_2}g^{\mu_3\mu_1}\bigg]$}}
\end{picture}
\begin{picture}(400,200)(0,0)
\SetColor{Black}
\Photon(80,110)(120,90){2}{5}\Text(70,115)[]{$\mu_2$}\Text(100,110)[]{$p_2$}
\ArrowLine(102,98)(97,103)
\Photon(120,90)(160,110){2}{5}\Text(170,115)[]{$\mu_4$}\Text(140,110)[]{$p_4$}
\ArrowLine(138,98)(143,103)
\Photon(80,70)(120,90){2}{5}\Text(70,65)[]{$\mu_1$}\Text(100,70)[]{$p_2$}
\ArrowLine(102,82)(97,77)
\Photon(120,90)(160,70){2}{5}\Text(170,65)[]{$\mu_2$}\Text(140,70)[]{$p_2$}
\ArrowLine(138,82)(143,77)
\Text(300,160)[]{=\quad{$-4ig^2\bigg[(g^{\mu_1\mu_2}g^{\mu_2\mu_4}-g^{\mu_1\mu_4}g^{\mu_2\mu_3})$}}
\Text(300,130)[]{\quad\quad{$\times\sin\Big({1\over2}p_1\tilde
    p_2\Big)\sin\Big({1\over2}p_3\tilde p_4 \Big)$}}
\Text(300,100)[]{\quad\quad{$+(g^{\mu_1\mu_4}g^{\mu_2\mu_3}-g^{\mu_1\mu_2}g^{\mu_3\mu_4})$}}
\Text(300,70)[]{\quad\quad{$\times\sin\Big({1\over2}p_3\tilde
    p_1\Big)\sin\Big({1\over2}p_2\tilde p_4 \Big)$}}
\Text(300,40)[]{\quad\quad{$+(g^{\mu_1\mu_2}g^{\mu_3\mu_4}-g^{\mu_1\mu_3}g^{\mu_2\mu_4})$}}
\Text(300,10)[]{\quad\quad{$\times\sin\Big({1\over2}p_1\tilde
    p_4\Big)\sin\Big({1\over2}p_2\tilde p_3 \Big)\bigg]$}}
\end{picture}
\begin{picture}(300,80)(0,0)
\SetColor{Black}
\DashArrowLine(50,60)(90,50){3}\Text(70,45)[]{$p_i$}
\DashArrowLine(90,50)(130,60){3}\Text(110,45)[]{$p_f$}
\Photon(90,20)(90,50){2}{3}\Text(100,30)[]{$p_\gamma$}
\Text(90,10)[]{$\mu$}\Text(240,50)[]{=\quad{$2igp^\mu_f
    \sin({i\over2}p_1\tilde p_f)$}}
\end{picture}\\
{\sl Vertex diagrams In NCQED}
\end{center}

%=======================================================%
% Soft Bremsstrahlung in the Noncommutative QED
%=======================================================%
\section{Soft Bremsstrahlung in the Noncommutative QED}
        \label{sobrem}
\begin{figure}[h]
  \begin{center}
    \includegraphics[scale=0.6]{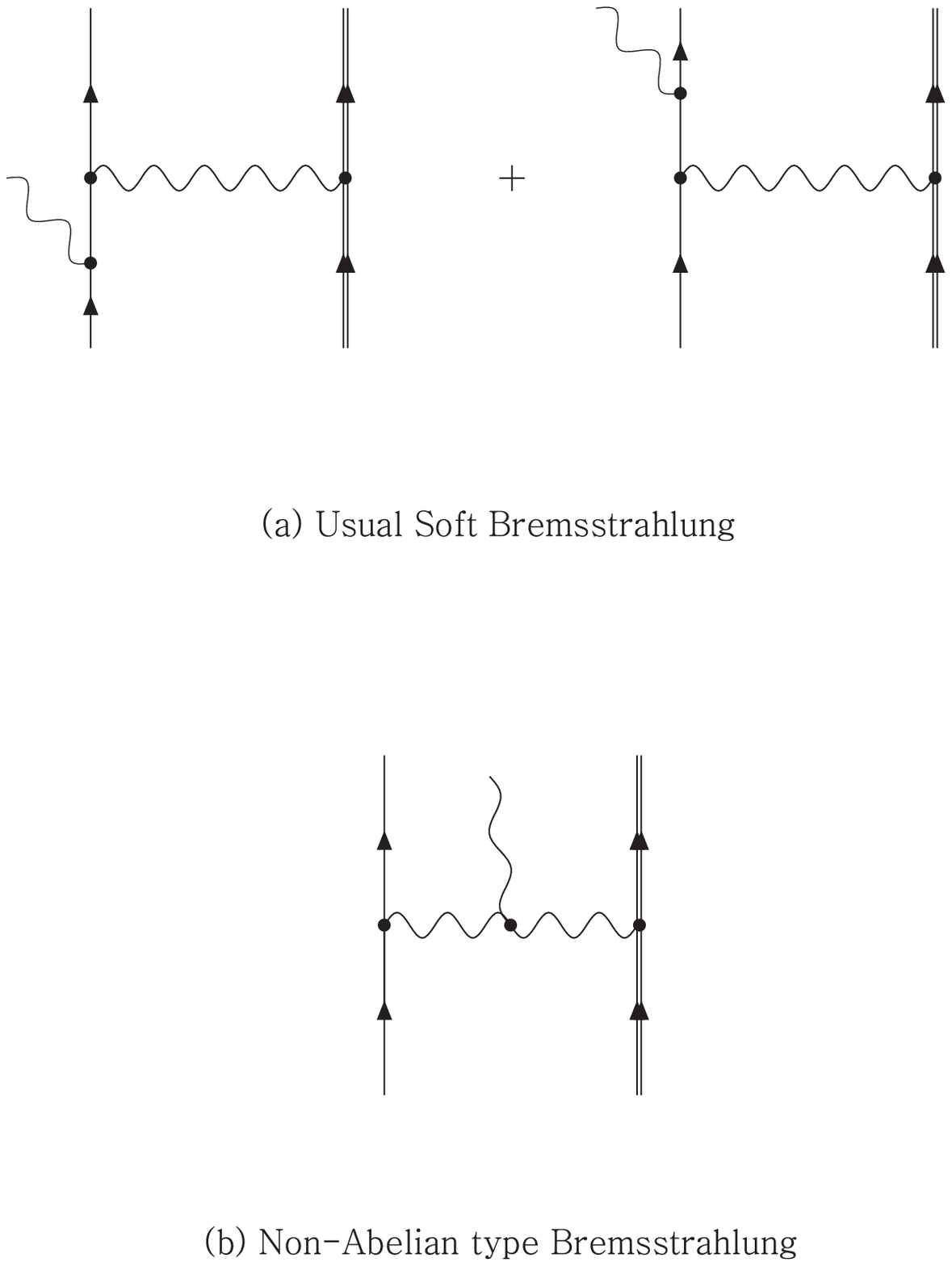}
    \caption{Soft Bremsstrahlung in the NCQED  (tree level)}
    \label{brem}
  \end{center}
\end{figure}

    \setcounter{equation}{0}
Now we study the radiative corrections by analyzing the bremsstrahlung
process. In addition to the ordinary diagrams in
(Fig.\ref{brem}a), we have an extra diagram (Fig.\ref{brem}b) due
to the new type of vertex in NCQED similar to those found in
non-Abelian gauge theories. We will evaluate the cross section
for all the three diagrams in (Fig.\ref{brem}) and investigate the
IR divergences for soft bremsstrahlung. \vs{3mm}
%==============================================%
%\subsection{Usual soft Bremsstrahlung}
%==============================================%
First, consider the diagrams (Fig.\ref{brem}a) for the usual soft
Bremsstrahlung. The amplitude $ M_{a} $ from  the diagram (a) is
\begin{eqnarray}
  iM_a& =&\phs
  \bar{u}(p_1')
  \Biggl[M_{0}(p_{1}',p-k)
  \frac{i({\dir{p_1}}-{\dir{k}}+m)}{(p_1-k)^{2}-m^2}
  \gamma^{\rho}\epsilon^{*}_{\rho}(k){e^{\frac{i}{2}g\theta{k}}}
  \nonumber
  \\
  &&+{\gamma^{\rho}}{\epsilon^{*}_{\rho}(k)}
  {e^{-\frac{i}{2}g\theta{k}}}
  \frac{{i}(\dir{p_{1}}+\dir{k}+m)}{(p_1+k)^{2}-m^2}
  M_{0}(p_{1}'+k,p)\Biggr]
  \bar{u}(p'_2)\frac{-ig_{\mu\nu}}{q^2}(-ie\gamma^{\nu})u(p_2)  .
\end{eqnarray}
In the above equation, $M_0$ is $-ie\gamma^{\mu}$ at the tree level and
${e^{\frac{i}{2}g\theta{k}}}$, ${e^{-\frac{i}{2}g\theta{k}}}$ are
reduced from the noncommutative phase factors.
 Since we are interested in the IR limit, we
assume the radiated photon being soft: $ |k| \ll |p'_{1}-p| $.
Then we can approximate
\begin{equation}
  M_{0}(p',p-k) \approx M_{0}(p'_{1}+k,p) \approx M_{0}(p',p)
\end{equation}
and can ignore $ \dir{k}$ in the numerators of the propagators.
The numerators can be further simplified with some Dirac algebra.
In the first term we have
\begin{eqnarray}
  (\dir{p}+m)\gamma^{\rho}\epsilon^{*}_{\rho}u(p)
  &=&\left[2p^{\rho}\epsilon^{*}_{\rho}
    +\gamma^{\rho}\epsilon^{*}_{\rho}(-\dir{p}+m)\right]u(p) \nonumber
  \\
  &=& 2p^{\rho}\epsilon^{*}_{\rho}u(p)
\end{eqnarray}

The denominators of the propagators are also simplified:
\begin{eqnarray}
  (p-k)^2-m^2=-2p\cdot k; & & (p'+k)^2-m^2=2p'\cdot k.
\end{eqnarray}
Hence in the soft-photon approximation, the amplitude becomes
\begin{eqnarray}
  iM_a=\bar{u}(p'_{1})[M_0(p'_1,p_1)]u(p_1)
  \left[e\cdot \left(\frac{p'_1\cdot \epsilon^{*}}{p'_1\cdot k}
      e^{\frac{i}{2}g\theta{k}}
      -\frac{p_1\cdot \epsilon^{*}}{p_1\cdot k}
      e^{-\frac{i}{2}g\theta{k}}\right) \right]
\end{eqnarray}
This is nothing but the amplitude for elastic scattering (without
bremsstrahlung)times a factor(in brackets) for the emission of
the photon \ct{peskin}.

%==============================================%
%\subsection{Non-Abelian type Bremsstrahlung}
%==============================================%
In the case of Non-abelian type Bremsstrahlung
(Fig.\ref{brem}b),  the amplitute $M_b$ becomes
\begin{eqnarray}
  iM_b&=&\phs\ul\frac{-ig_{\mu\alpha}}{{q'}^2}\frac{-g_{\beta\nu}}{q^2}
  \epsilon^*_\rho(k)\biggl[B\biggr]\ur
\end{eqnarray}
%where $M_o=-ie\gamma^\mu$ for the tree level
where $[B]\;$ is the phase factor for the three photon
vertex(Non-Abelian type) given by
\begin{eqnarray}
  \biggl[B\biggr]=-2e\sin\Bigl(\frac{k\cdot \ti{q'}}{2}\Bigr)
  \biggl[g^{\rho\alpha}(k-q)^\beta+g^{\alpha\beta}(q+q')^\rho+g^{\beta\rho}(-q'-k)^\alpha\biggr]
\end{eqnarray}
This is simplified as
\begin{eqnarray}
  iM_b&=&\phs\ul \frac{2e\sin\Bigl(\frac{k\cdot
      \ti{q'}}{2}\Bigr)}{(q-k)^2q^2}\epsilon^*_\rho \n \\
  &&\biggl[g^\rho_{\mu}(k-q)_\nu+g_{\mu\nu}(q+q')^\rho+g^\rho_{\nu}
  (-q'-k)_\mu\biggr]
  \times\ur
\end{eqnarray}
There is no IR divergence in this expression. That is, $M_b$ is
finite for the soft photon : $k\ll|p'_1-p_1|$.

%==============================================%
%\subsection{Cross section for the soft photon}
%==============================================%

The cross section for the Bremsstrahlung is  expressed in terms
of the elastic cross section by inserting an additional
phase-space integration for the photon variable $k$. Summing over
the two photon polarization states, we obtain
\begin{eqnarray}
  iM&=&iM_a+iM_b\n\\
  |M|^2&=&M_a^2+M_b^2+M_a^*M_b+M_aM_b^*
\end{eqnarray}
In this expression, only $M_a^2$ contribute to IR divergence.
Thus evaluating the cross section only for the usual QED diagram
(a), is enough for IR divergence purpose.
\begin{eqnarray}
  d\sigma(p\rightarrow p'+\gamma) & = & d\sigma(p\rightarrow
  p')\nonumber
  \\
  && \cdot \int \frac{d^3k}{(2\pi)^3}\frac{1}{2k}\sum_{
    \lambda=1,2}e^2 \left \vert \frac{p'\cdot \epsilon^{(\lambda)}}{p'\cdot
      k}{e^{\frac{i}{2}g\theta{k}}}
    -\frac{p\cdot \epsilon^{(\lambda)}}{p\cdot k}{e^{-\frac{i}{2}g\theta{k}}}\right
    \vert^2 \n \\
\end{eqnarray}
The differential probability of radiating a photon with momentum
$k$, given by an election scattered from $p_1$ to $p'_1$,
reads
\begin{eqnarray}\l{pr}
  d(prob)=\frac{d^3k}{(2\pi)^3}\sum_{\scriptstyle\lambda}\frac{e^2}{2k}
  \left|\epsilon_\lambda \cdot \left(\frac{p'_1{e^{\frac{i}{2}g\theta{k}}}}{p'_1\cdot
        k}-\frac{p{e^{-\frac{i}{2}g\theta{k}}}}{p\cdot k}\right)\right|^2
\end{eqnarray}
Multiplying by the photon energy $k$ will give the radiated
energy.

The equation({\ref{pr}}) is an expression not for the expected
number of photon radiated, but for the probability of radiating a
single photon. The problem becomes worse if we integrate over
photon momentum. In order for the soft-photon approximation to be available,
the integration upper limit must be restricted. So
we will integrate only up to the energy scale
where the soft-photon approximation is broken; a reasonable estimate
for this energy is $|q|=|p'_1-p_1|$. The integral is therefore
\begin{eqnarray}
\l{I}
  \textrm{Total \; probability} \approx  \frac{\alpha}{\pi} \int_0^{|q|}dk
  \frac{1}{k}I_{N.C}
\end{eqnarray}
where $I_{N.C}$ denote essentially the differential intensity
$d$(Energy)/$dk$ for NCQED. We find that the radiative energy at
low frequencies($k\rightarrow0$) is given by
\begin{eqnarray}
  I_{N.C}&=& I_{C}\cdot \cos(q\cdot\ti{k})\n \\
  &\approx&2\log\left(\frac{-q2}{m^2}\right)\cdot \cos(q\cdot
  \ti{k})
\end{eqnarray}
where $I_c$ represents the differential intensity for the
commutative QED. We can regularize the integral in (\ref{I}) by
introducing the very small photon mass $\mu$. This mass would
then provide a lower cutoff for the integration over the soft
photon momentum,
\begin{eqnarray} \l{so}
    d\sigma(p\rightarrow
    p'+\gamma)&=&d\sigma(p\rightarrow p') \cdot \frac{\alpha}{\pi}
    \int_\mu^{|q|}dk
    \frac{1}{k}I_{C}\cdot\cos(q\cdot
    \ti{k})\n\\
    &\approx&d\sigma(p\rightarrow p') \cdot\frac{\alpha}{\pi}
    \log\left(\frac{-q^2}{\mu^2}\right)
    \log\left(\frac{-q^2}{m^2}\right)
\end{eqnarray}

\section{Vacuum Polarization in the Noncommutative QED}
        \label{vacpol}
    \setcounter{equation}{0}

\begin{figure}[h]
    \begin{center}
      \includegraphics[scale=.8]{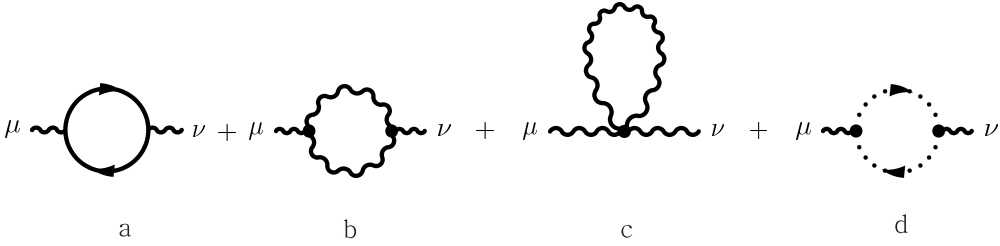}
      \caption{Vacuum polarization in the NCQED} \label{vacd}
    \end{center}
\end{figure}

We consider the 2-point photon self energy diagrams.  The
contributions are from loops involving fermion, scalars and gauge
bosons (Fig.\ref{vacd}).
Applying the NC Feynman rules, we find the matrix element $M$ of
the photon self energy diagrams.
\begin{eqnarray}
    iM&=&iM^a+iM^b+iM^c+iM^d\n\\
    &=&\bar{u}(p'_1)(-ie\gamma^{\mu}) u(p_1)
    \frac{-i}{q^2}
    \Bigl[i\Pi\Bigr]
    \frac{-i}{q^2}
    \bar{u}(p'_2)(-ie\gamma^\nu)u(p_2)
\end{eqnarray}
where,
$i\Pi=i\Pi^{a}_{\mu\nu}+\Pi^{b}_{\mu\nu}+\Pi^{c}_{\mu\nu}+\Pi^{d}_{\mu\nu}$
%    && \textrm{where,}\quad\quad i\Pi=i\Pi^{a}_{\mu\nu}+\Pi^{b}_{\mu\nu}+\Pi^{c}_{\mu\nu}+\Pi^{d}_{\mu\nu}
with
% ${i\Pi^{a}_{\mu\nu}, i\Pi^{b}_{\mu\nu}, i\Pi^{c}_{\mu\nu},
%i\Pi^{d}_{\mu\nu}}$ is,
\begin{eqnarray}
    i\Pi^a_{\mu\nu}&=&-e^2\int\frac{d^4k}{(2\pi)^4}
    \frac{\textmd{tr}[\gamma_\mu(\dir{k}+m)\gamma_{\nu}(\dir{k}-\dir{q}+m)]}
    {(k^2-m^2)((k-q)^2-m^2)}\n\\
    i\Pi^b_{\mu\nu}
    &=&-4e^2\int\frac{{d^4}k}{(2\pi)^4}
    \frac{\sin^2\Bigl(\frac{k\cdot \ti{q}}{2}\Bigr){Q_{\mu\nu}}}
    {(k-q)^2 {k^2}} \n\\
    &&Q_{\mu\nu}=g_{\mu\nu}(-5q^2-2k^2+2q\cdot k)+5(k_\mu q_\nu+
    k_\nu q_\mu) + 2q_\mu q_\nu -10k_\mu k_\nu
\end{eqnarray}
\begin{eqnarray}
    i\Pi^c_{\mu\nu}&=&-12e^2\intk
    \frac{g_{\mu\nu}\sin^2\Bigl(q\cdot \ti{k}\Bigr)}
    {k^2}\n\\
    i\Pi^d_{\mu\nu}&=&-4e^2\intk \frac{(k-q)_\mu
    \sin^2\Bigl(\frac{q\cdot \ti{k}}{2}\Bigr)}
    {(k-q)^2 k^2}
\end{eqnarray}

 All the UV divergences can be subtracted away by the same local
counterterms as those in the ordinary QED {\cite{hya}}. Our main
concern is  the IR divergences of these diagrams. As
$q\rightarrow0$, all diagrams are finite. Hence the structure
of the  IR divergences is the same as those in the ordinary QED.

\section{Vertex structure at the one loop level in NCQED}
        \label{verfun}
    \setcounter{equation}{0}

\begin{figure}[h]
  \begin{center}
    \includegraphics[scale=0.9]{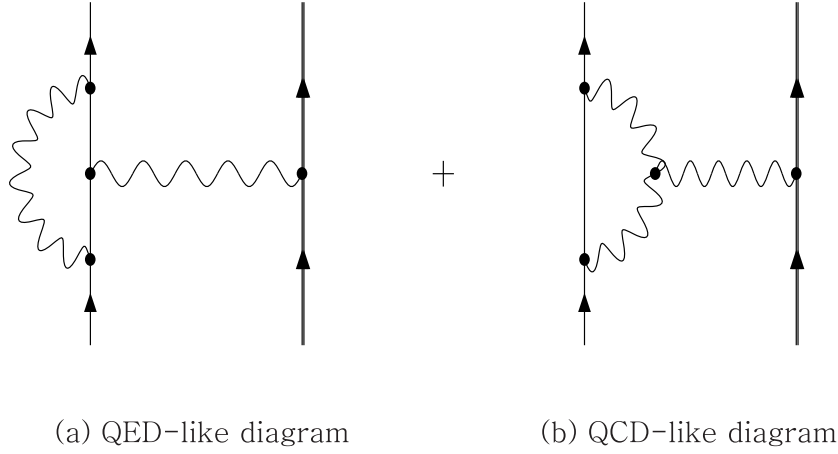}
    \caption{One loop correction to $\psi\bar{\psi}A_\mu$ vertex}
    \label{vtf}
  \end{center}
\end{figure}

  In this section we perform explicitly the calculation of the vertex
function for the photon-electron at the one loop level. Due to the
three photon vertices in NCQED, the radiative corrections to the
electron-photon vertex come from the two diagrams of Fig
\ref{vtf}. \vs{5mm}
%===================================================
%==============================================%
%\subsection{QED-like diagram}
%==============================================%
The invariant matrix element $M$ is given by
\begin{equation}
iM=iM_1 + iM_2.
\end{equation}
The $M_1$ for the QED-like diagram (Fig3.a) is derived as
%We know
%formulae for cross sections in terms of the invariant matrix
%element $M$,  for various processes in various interesting field
%theories. we can compute $M$ using Feynman diagrams.
\begin{eqnarray}
    iM_1&=&\phs \bar{u}(p'_1)
    \Bigl(-ie\Lambda^\mu_1(p'_1,p_1)\Bigr)u(p_1)
   \times\frac{-ig_{\mu\nu}}{q^2}\bar{u}(p'_2)(-ie\gamma^\nu)u(p_2)
\end{eqnarray}
%Applying the Feynman rules,
where
$$\Lambda^\mu_1 (p'_1,p_1)=\gamma^\mu +\Gamma^\mu_1 (p'_1,p_1)$$
with
\begin{eqnarray}
    \Gamma^\mu_1 (p',p)&=&i(-ie)^2\intk \ekq
    \frac{\gamma^\rho}{k^2-\mu^2+i\epsilon}
    \frac{\dir{p'}-\dir{k}+m}{(p'-k)^2-m^2+i\epsilon}\gamma^\mu \n \\
    &\times& \frac{\dir{p}-\dir{k}+m}{(p-k)^2-m^2+i\epsilon}
    \gamma_\rho
\end{eqnarray}
Here we have added the fictitious photon mass $\mu$ as an IR regulator
of the integration.

With the mass shell condition, the numerator in the above
expression may be written as
\begin{equation}
    4\left\{\gamma^\mu\left[(p'-k)\cdot(p-k)-\frac{k^2}{2}\right]
    +(p'+p-k)^\mu \dir{k} -m k^\mu\right\}
\end{equation}
We use the following Schwinger parameter representation of the
propagators.
\begin{equation}
    \frac{i}{p^2-m^2+i\epsilon}
    = \int_0^\infty d\alpha e^{i\alpha(p^2-m^2+i\epsilon)}
\end{equation}
The integral converges at the upper limit owing to the presence
of $i\epsilon$. We  introduce the following auxiliary integral as
the generating function $Z$\ct{itz}.
%for the various terms of integration.
\begin{eqnarray}
    Z\equiv\intk e^{ik\cdot
    (z-\tilde{q})}
    \frac{1}{(k^2-\mu^2+i\epsilon)(k^{2}-2p'\cdot k+i\epsilon)
    (k^2-2p\cdot k+i\epsilon)} \n \\
    =
    \frac{1}{(4\pi)^2}
    \int_0^\infty
    \frac{d\al{1}d\al{2}d\al{3}}{(\al{1}+\al{2}+\al{3})^2}
    exp\left\{-i\left[\al{1}\mu^2+
    \frac{\left(\frac{(z-\tilde{q})}{2}-\al{2}p'-\al{3}p\right)^2}
    {\al{1}+\al{2}+\al{3}}\right]\right\}\n \\
\end{eqnarray}
Then the integration with powers of $k$ in the numerator can be
obtained by differentiating the above generating function with
respect to $z$.
%The introduction of the factor $e^{ik\cdot z}$ in
%the integrand allows us to obtain the required expression in the
%numerator of $\Gamma^\mu$ by differentiation({\ref{dif}}).

%\begin{eqnarray} \l{dif}
%    \Gamma^\mu_1 (p',p)
%    \left\{
%    \begin{array} {l}
%    {k^\mu={\frac{1}{i}\frac{\partial Z}{\partial
%    z_\mu}}\mid_{z\rightarrow 0}
%    ={-\frac{(\frac{(z-\tilde{q})}{2}-\al{2}p'-\al{3}p)^\mu}{\al{1}+\al{2}+\al{3}}}
%    Z\mid_{z\rightarrow 0}}
%    \\
%    {\dir{k}=k^\mu\gamma_\mu=-\frac
%    {(\frac{z-\not{\tilde{q}}}{2}-\al{2}\not{p'}-\al{3}\not{p})}
%    {\al{1}+\al{2}+\al{3}}
%    Z\mid_{z\rightarrow 0}}
%    \end{array} \right.
%\end{eqnarray}
For instance, the integration with $k^\mu$ in the numerator is
given by
$${\frac{1}{i}\frac{\partial Z}{\partial
    z_\mu}}\mid_{z\rightarrow 0}
    ={-\frac{(\frac{(z-\tilde{q})}{2}-\al{2}p'-\al{3}p)^\mu}{\al{1}+\al{2}+\al{3}}}
    Z\mid_{z\rightarrow 0}$$
%<<<<<<<<<<<<<<<<<<<<<<<<<REVISED>>>>>>>>>>>>>>>>>>>>>>>>>>>>>>>%
%\textit{For instance,  $k^\mu$ in the numerator is obtained by
%differentiation of generating function $Z$ by vabiable $z_\mu$}
%<<<<<<<<<<<<<<<<<<<<<<<<<REVISED>>>>>>>>>>>>>>>>>>>>>>>>>>>>>>>%
After symmetrization in $\al{1}$ and $\al{2}$ we obtain the
following representation
\begin{eqnarray}
    \Gamma^\mu_{1} (p'_1,p_1)&=&
    \frac{\alpha}{i\pi}\phs\int_0^\infty\frac{d\al{1}d\al{2}d\al{3}}{(\al{1}+\al{2}+\al{3})^3}
    \Biggl
    [\gamma^\mu
    \biggl\{(\al{1}+\al{2}+\al{3})(p'_1\cdot p_1)\n \\
    &&-(\al{2}+\al{3})\frac{(p'_1+p_1)^2}{2}
    +p_1\cdot\tilde{p'_1}
    +\frac{i}{2}
    +\frac{\left[m^2(\al{2}+\al{3})^2-\al{2}\al{3}q^2\right]}
    {2(\al{1}+\al{2}+\al{3})}\n \\
    &&-\frac{(\al{2}+\al{3})(p'_1+p_1)\cdot\tilde{q}}{4(\abc)}
    +\frac{\frac{\tilde{q}^2}{4}}{(2\abc)}
    \biggr\}\Biggr]\n \\
    &&\times\exp\frac{i}{\abc}\biggl[\al{2}\cdot\al{3}q^2-(\al{2}+\al{3})^2m^2
    \n \\
    && -\al{1}(\abc)\mu^2+\frac{\al{2}+\al{3}}{2}(p'_1+p_1)
    \cdot\tilde{q}-\frac{\tilde{q}^2}{4}\biggr]
\end{eqnarray}
where we have set $p'_1-p_1=q$.

We observe that $ q^2 \leq 0$ if $p_1$ and $p'_1$ lie on the mass
shell. Using the identity $1=\int_0^\infty
d\rho\delta(\rho-\al{1}-\al{2}-\al{3})$,
changing the variables $\al{i}\rightarrow \rho \al{i}$ and
inserting UV regulator
$\exp(\frac{i}{\Lambda-\frac{\tilde{q}^2}{4}})$,  we obtain the
following after the Wick rotation $\al{i}\rightarrow-i\al{i}$ :
\begin{eqnarray}
  {\Gamma^{\mu}_{1}(p_1^{'},p_1)} &=& -\frac{\alpha}{\pi} \phs \int_0^1
    {d\alpha_{1}d\alpha_{2}d\alpha_{3}}\delta(1-\Sigma\alpha_{i})\times{e^{-b'}}\nonumber
    \\
    && \times \Biggl[\gamma^{\mu} \biggl(G^{(1)}_{a}\int_0^\infty {d\rho}e^{-({a\rho}+b/{\rho})}
    +(G^{(1)}_{b}+G^{(1)}_{c})\int_0^\infty
    \frac{d\rho}{\rho}e^{-({a\rho}+b/{\rho})}
     \n \\
    &&+G^{(1)}_{b^{'}}\int_0^\infty
    \frac{d\rho}{\rho^{2}}e^{-({a\rho}+b/{\rho})}\biggr)
    + \Bigl\{ H^{\mu,(1)}_{a}\int_0^\infty {d\rho}e^{-({a\rho}+b/{\rho})}
    \n \\
    &&+H^{\mu,(1)}_{b}\int_0^\infty \frac{d\rho}{\rho}e^{-({a\rho}+b/{\rho})}
    +H^{\mu,(1)}_{b^{'}}\int_0^\infty \frac{d\rho}{\rho^{2}}e^{-({a\rho}+b/{\rho})}\Bigr\}\Biggr]
    \n \\
    &&-(p^{'}=p, q=0)
\end{eqnarray}
where
\begin{eqnarray}
  \label{2}
  G^{(1)}_{a}&=&p_1\cdot{p_1}^{'}-\frac{(\alpha_{2}+\alpha_{3})(p_1^{'}+p_1)^{2}}{2}+
  \frac{1}{2}\left\{m^{2}(\alpha_{2}+\alpha_{3})^{2}-\alpha_{2}\alpha_{3}q^{2}\right\}
  \n \\
  G^{(1)}_{b}&=&i\left[\frac{1}{2}(p_1^{'}+p_1)
    \cdot\widetilde{q}-\frac{1}{4}(\alpha_{2}+\alpha_{3})(p_1^{'}+p_1)\cdot\widetilde{q}\right],
  \;\;\;
  G^{(1)}_{b^{'}}=-\frac{1}{8}\widetilde{q}^{2}
  \n \\
  G^{(1)}_{c}&=& - \frac{1}{2},
  \;\;\;
  H^{\mu,(1)}_{a}=\frac{m}{2}(p_1^{'}+p_1)^{\mu}\alpha_{1}(\alpha_{2}+\alpha_{3})
  \n \\
  H^{\mu,(1)}_{b}&=&i\left[\frac{m}{2}(1+\alpha_{2}+\alpha_{3})\widetilde{q}^{\mu}
    +\frac{1}{4}(\alpha_{2}+\alpha_{3}-2)(p_1^{'}+p_1)^{\mu}\widetilde{q}^{\sigma}\gamma_{\sigma}\right]
  \n \\
  H^{\mu,(1)}_{b^{'}}&=&\frac{1}{4}\widetilde{q}^{\mu}\widetilde{q}^{\nu}\gamma_{\nu}
\end{eqnarray}
and
\begin{eqnarray}
  \label{3}
  a&=&-\alpha_{2}\alpha_{3}q^{2}+(\alpha_{2}+\alpha_{3})^{2}m^{2}+\alpha_{1}\mu^{2}
  \n \\
  b &=& \frac{1}{\Lambda^{2}_{eff}}(={\Lambda^{-2}+\tilde{b}})\;\; (
  \tilde{b}=-\frac{1}{4}\widetilde{q}^{2})
  \n \\
  b' &=&\frac{i}{2}(\alpha_{2}+\alpha_{3})(p_1^{'}+p_1)\cdot\widetilde{q}
\end{eqnarray}
%The subscript $a$ denotes the ordinary case while $b,b',c$ the
%noncommutative effects.
The Lorentz structure of the terms with $G$ are from the
$\gamma^\mu$ while those of $H^\mu$ not. The constant $a$ and the
terms with subscripts $a,c$ are those appearing in the ordinary
QED. On the other hands, the constants $b,b'$ and the terms with
subscripts $b,b'$ are the additional terms coming from the
noncommutative effects. Hence, if we take the noncommutative
parameter $\theta$ going to zero limit, then the terms with
subscripts $b,b'$ go to zero and the results are those of the
ordinary QED.

The integrands with subscripts $b$ and $c$ are proportional to
one over $\rho$ while those with $b'$ are one over $\rho^2$.
%<<<<<<<<<<<<<<<<<<<<<<<<<REVISED>>>>>>>>>>>>>>>>>>>>>>>>>>>>>>>%
%\textit{Where the $G$ includes the $\gamma^\mu$ while $H$ not. The
%subscript $a,c$ denotes the ordinary case while $b,b'$ the
%noncommutative effects. In particular the $b, c$ are the one over
%$\rho$ integral form and $b'$ terms are the one over $\rho^2$
%integral. If the terms including $b,b'$ go to zero, the rest form
%are same as in ordinary case.}
%<<<<<<<<<<<<<<<<<<<<<<<<<REVISED>>>>>>>>>>>>>>>>>>>>>>>>>>>>>>>%
If we evaluate $\rho$ integral, we get Bessel functions of the
following form,
\begin{eqnarray}
\int_0^\infty d\rho e^{-D}&=&\int_0^\infty d\rho
e^{-(a\rho+{b\over\rho})}={1\over a}\int_0^\infty d\rho
e^{-(\rho+{ab\over\rho})}\nonumber\\&=&{1\over a}\Big\{2\sqrt{ab}
K[1,2\sqrt{ab}]\Big\}\nonumber\\
\int_0^\infty {d\rho\over\rho}e^{-D}&=&\int_0^\infty
{d\rho\over\rho}e^{-(\rho+{ab\over\rho})}=
\Big\{2K[0,\sqrt{ab}]\Big\}\nonumber\\
\int_0^\infty {d\rho\over\rho^2}e^{-D}&=&a\int_0^\infty
{d\rho\over\rho^2}e^{-(\rho+{ab\over\rho})}=a\left\{{2\over\sqrt{ab}}K[1,2\sqrt{ab}]\right\}\nonumber\\
\end{eqnarray}
where $K_0$, $K_1$ are the modified Bessel function.

With the above results of the integral over $\rho$, we get
\begin{eqnarray}
  \Gamma^\mu_1(p'_1,p_1)=-\frac{\alpha}{\pi}\phs \inta
    e^{-i(\al{2}+\al{3})(p'_1+p_1)\cdot \tilde{q}}\n \\
    \Biggl (
    \gamma^\mu
    \biggl\{G^{(1)}_a\frac{2\sqrt{ab}K_1(a\sqrt{ab})}{a}
    +(G^{(1)}_b+G^{(1)}_c)2K_0(2\sqrt{ab})+G^{(1)}_{b'}\frac{aK_1(2\sqrt{ab})}{\sqrt{ab}}\biggr\}
    \n \\
    +H^{\mu,(1)}_a \frac{2\sqrt{ab}K_1(a\sqrt{ab})}{a}
    +H^{\mu,(1)}_b 2K_0(2\sqrt{ab})
    +H^{\mu,(1)}_{b'} \frac{aK_1(2\sqrt{ab})}{\sqrt{ab}}
    \Biggr)\n \\
\end{eqnarray}

%==============================================%
%\subsection{QCD-like diagram}
%==============================================%
We now evaluate  the matrix element $M_2$ for the second QCD-like
diagram(Figure3 (b)). It becomes
\begin{equation}
    iM_2=\bar{u}(p'_1)
    \Bigl(-ie\Lambda^\mu_2(p'_1,p_1)\Bigr)u(p_1)
    \frac{-ig_{\mu\nu}}{q^2}\bar{u}(p'_2)(-ie\gamma^\nu) e^{\frac{i}{2}(p_2\cdot \ti{p'_2})} u(p_2)
\end{equation}
where $\Lambda^\mu_2 (p'_1,p_1)=\gamma^\mu
+\Gamma^\mu_2(p'_1,p_1)$, and
%The analytic expression of which reads as
\begin{eqnarray}
    \Gamma^\mu_2 &=&
    \intk (-ie\gamma^\nu) e^{\frac{i}{2}(k\cdot \tilde{p'_1})}
    \frac{i(\dir{k}+m)}{k^2-m^2}(-ie\gamma_\rho)
    e^{\frac{i}{2} p_1\cdot (\tilde{p_1}-\tilde{k})}
    \frac{-i}{(p'_1-k)^2-\mu^2} \n \\
    &&\times\frac{-i}{(p_1-k)^2-\mu^2} \Biggl\{(-2e)\sin{\Bigl(\frac{1}{2}(p'_1-k)\cdot (-\ti{p}+\ti{k})\Bigr)}
    \n
    \\
    &&\times\biggl[g^{\nu\rho} (p'_1+p_1-2k)^\mu + g^{\rho\mu} (-p_1+k+q)^\nu
    +g^{\mu\nu} (-q-p'_1+k)^\rho \biggr]\Biggr\} \n\\
    &=& i(-ie)^2 \intk \frac{\phs (1-e^{-iq\cdot
    \ti{k}}e^{-ip\cdot \ti{p'}})}
    {[k^2-m^2][(p_1-k)^2-\mu^2][(p'_1-k)^2-\mu^2]} \n \\
    &&\times
    \biggl\{\gamma_nu(\dir{k}+m)\gamma_\rho
    \Bigl[g^{\nu\rho} (2k-p'_1-p_1)^\mu + g^{\rho\mu}
    (2p_1-p'_1-k)^\nu\n \\
    &&+g^{\mu\nu}
    (2p'_1-p_1-k)^\rho \Bigr]
    \biggr\}
\end{eqnarray}
with $p'_1-p_1=q$. Using the mass shell condition and gamma matrix
algebra, the numerator can be written as
\begin{equation}
    8mk_\mu - 2\dir{k}(p'+p+2k)_\mu+2\gamma_\mu(2p'.k+2p.k-k^2-3m^2)\
\end{equation}
We rewrite the matrix element using the  Schwinger parameter
representation for  the propagators as before.
%%We use the following Schwinger parameter representation of the
%propagators like previous section.
%\begin{equation}
%    \frac{i}{p^2-m^2+i\epsilon}
%    = \int_0^\infty d\alpha e^{i\alpha(p^2-m^2+i\epsilon)}
%\end{equation}

Here also the introduction of the auxiliary integral as the
generating function $Z'$ for the various integration is very
helpful.
\begin{eqnarray}
    Z'&=&\intk
    \frac{e^{ik\cdot (z+\ti{q})}}{(k^2-m^2)((p'_1-k)^2-\mu^2)((p_1-k)^2-\mu^2)} \n \\
    &=&
    \frac{1}{4\pi^2} \int_0^\infty
    \frac{d\al{1}d\al{2}d\al{3}}{(\abc)^2} \n \\
    &&\times\exp\Biggl\{\frac{-i}{(\abc)}\biggl[\frac{z^2}{4}-z\cdot
    (\al{2}p'_1+\al{3}p_1-\frac{\tilde{q}}{2})\n \\
    &&-(\al{2}+\al{3})p_1\cdot \ti{q}
    +m^2(\al{1}-\al{2}-\al{3})(\abc)\n \\
    &&+\mu^2(\al{2}+\al{3})(\abc)
    +m^2(\al{2}+\al{3})^2-\al{2}\al{3}q^2-b\biggr]\Biggr\}\n \\
\end{eqnarray}
where we have inserted a UV regulator,
$b=\frac{1}{\Lambda^{2}_{eff}}(={\Lambda^{-2}+\tilde{b}}) \;\;(
\tilde{b}=-\frac{1}{4}\widetilde{q}^{2})$.

The integration with  various powers of $k$ in the numerator  is
produced through the derivatives over the $Z'$. Inserting the
identity $1=\int_0^\infty
d\rho\delta(\rho-\al{1}-\al{2}-\al{3})\;$, rescaling
$\al{i} \rightarrow \rho \al{i}\;$, and Wick rotation, we finally
obtain
\begin{eqnarray}
    \Gamma^\mu_2&=&
    \frac{-\alpha}{\pi}\phs \inta e^{i(\al{2}+\al{3})p_1\cdot
    \ti{q}}e^{-ip\cdot \ti{p'}}\n \\
    &\times&
     \Biggl[\gamma^{\mu} \biggl(G^{(2)}_{a}\int_0^\infty {d\rho}e^{-({a\rho}+b/{\rho})}
    +(G^{(2)}_{b}+G^{(2)}_{c})\int_0^\infty \frac{d\rho}{\rho}e^{-({a\rho}+b/{\rho})}
     \n \\
    &&+G^{(2)}_{b^{'}}\int_0^\infty
    \frac{d\rho}{\rho^{2}}e^{-({a\rho}+b/{\rho})}\biggr)
    + \Bigl\{ H^{\mu,(2)}_{a}\int_0^\infty {d\rho}e^{-({a\rho}+b/{\rho})}
    \n \\
    &&+H^{\mu,(2)}_{b}\int_0^\infty \frac{d\rho}{\rho}e^{-({a\rho}+b/{\rho})}
    +H^{\mu,(2)}_{b^{'}}\int_0^\infty \frac{d\rho}{\rho^{2}}e^{-({a\rho}+b/{\rho})}\Bigr\}\Biggr]
    \n \\
    &+&\frac{-\alpha}{\pi}\phs \inta \n \\
    &&\times
    \Biggl[\gamma^{\mu} \biggl(G^{(2)}_{a}\int_0^\infty {d\rho}e^{-({a\rho}+b'/{\rho})}
    +G^{(2)}_{c}\int_0^\infty \frac{d\rho}{\rho}e^{-({a\rho}+b'/{\rho})}\biggr)
     \n \\
     && + \Bigl\{ H^{\mu,(2)}_{a}\int_0^\infty {d\rho}e^{-({a\rho}+b'/{\rho})}
    \Bigr\}\Biggr]
    -(p^{'}=p, q=0)
\end{eqnarray}
where,
\begin{eqnarray}
    G^{(2)}_a &=&
    \frac{1}{2}\Bigl[(\al{2}+\al{3})(p'_1+p_1)^2-3m^2-m2(\al{2}+\al{3})^2+\al{2}\al{3}q^2\Bigr]\n
    \\
    G^{(2)}_b &=& \frac{-i}{2}p_1\cdot \ti{q}(2-\al{2}-\al{3}),
    \;\;\;\;
    G^{(2)}_{b'} = \frac{\ti{q}^2}{8}, \;\;\;\;
    G^{(2)}_c = \frac{3}{2}\n\\
    H^{\mu,(2)}_a &=& \frac{m}{2}(p'_1+p_1)^{\mu}\al{1}(\al{2}+\al{3})\n
    \\
    H^{\mu,(2)}_b &=& \frac{-i}{2}(2-\al{2}-\al{3})\ti{q}^\mu
    +\frac{i}{4}(1+\al{2}+\al{3})(p'_1+p_1)^{\mu}\gamma\cdot
    \ti{q}\n \\
    H^{\mu,(2)}_{b'} &=& \frac{\gamma\cdot \ti{q}}{4}\ti{q}^\mu \n \\
\end{eqnarray}
with
\begin{eqnarray}
    a&=&m^2(\al{1}-\al{2}-\al{3})+\mu^2(\al{2}+\al{3})+m^2(\al{2}+\al{3})^2-\al{2}\al{3}q^2\n\\
    b&=&{\Lambda^{2}_{eff}(=\Lambda^{-2}+\tilde{b})} \;\;
    (\tilde{b}=-\frac{1}{4}\widetilde{q}^{2})
\end{eqnarray}
Here also the constant $b$ and terms $G$ or $H$ with the
subscripts $b, b'$  are the additional quantities coming from the
noncommutativity.  In the limit of ordinary QED, those values
with indices $b$ and $b'$ become zero and we get the same
result as that in the ordinary QED.

Integration over $\rho$ leads to
\begin{eqnarray}
    \Gamma^\mu_2&=&\frac{-\alpha}{\pi}\phs \inta e^{i(\al{2}+\al{3})p_1\cdot
    \ti{q}}e^{-ip\cdot \ti{p'}}\n \\
    &&\times \Biggl[\gamma^\mu \biggl\{G^{(2)}_a\frac{2\sqrt{ab}K_1(a\sqrt{ab})}{a}
    +(G^{(2)}_b+G^{(2)}_c)2K_0(2\sqrt{ab})+G^{(2)}_{b'}\frac{aK_1(2\sqrt{ab})}{\sqrt{ab}}\biggr\}\n \\
     &&\quad\quad\quad
     +H^{\mu,(2)}_a \frac{2\sqrt{ab}K_1(a\sqrt{ab})}{a}+H^{\mu,(2)}_b 2K_0(2\sqrt{ab})
    +H^{\mu,(2)}_{b'} \frac{aK_1(2\sqrt{ab})}{\sqrt{ab}}
    \Biggr]\n \\
    &+&\frac{-\alpha}{\pi}\phs \inta \n \\
    &&\times \Biggl[\gamma^\mu \biggl\{G^{(2)}_a\frac{2\sqrt{ab'}K_1(a\sqrt{ab'})}{a}
    +G^{(2)}_c 2K_0(2\sqrt{ab'})\biggr\}+H^{\mu,(2)}_a \frac{2\sqrt{ab'}K_1(a\sqrt{ab'})}{a}
    \Biggr]
\end{eqnarray}

%==============================================%
%\subsection{Renormalization}
%==============================================%
We want analyze the divergences in NCQED and compare with those
in the ordinary QED, where the logarithmic UV divergences are
renormalized and the problem of IR divergences , after
regularized by the soft photon mass $\mu$, are cancelled between
the bremstrahlung and the radiative loop corrections.
%with was fixed by the photon a finite mass, .

In section \ref{irdiv}, we will analyze in detail the IR
divergence of NCQED.

The vertex functions contain $K_0$ and $K_1$, and both of the
functions contain either UV regulator $\frac{1}{\Lambda_{eff}^2}$
or $\frac{1}{\Lambda^2}$ in their arguments. As the high energy
limit $\Lambda^2 \to \infty$, or $b\rightarrow 0$, we find that
all terms containing $K_1$ are finite, but a
logarithmic divergence appears in $K_0$. Since the noncommutative QED
was shown to be renormalizable up to the one loop level by adding
the relevant counter terms, we can safely drop the singular parts
in $K_0$, keeping only the finite parts.

\begin{eqnarray} \l{ren0}
    \Gamma^\mu_{(R)}&=&\Gamma^\mu_{1(R)}+\Gamma^\mu_{2(R)}\n\\
    &=&\frac{-\alpha\phs}{\pi}  \inta \n\\
    &&\Biggl[
    \frac{(\gamma^\mu G^{(1)}_a + H^{\mu,(1)}_a)e^{-i(\al{2}+\al{3}) p_1\cdot
    \ti{q}}}{a}+\frac{(\gamma^\mu G^{(2)}_a + H^{\mu,(2)}_a)(1-e^{i(\al{2}+\al{3}) p_1\cdot
    \ti{q}})e^{-ip_1\cdot p'_1}}{a'}\n\\
    &&\quad-2\gamma_E \biggl\{(\gamma^\mu G^{(1)}_b+\gamma^\mu G^{(1)}_c +H^{\mu,(1)}_b)
    e^{-i(\al{2}+\al{3}) p_1\cdot
    \ti{q}} +\gamma^\mu G^{(2)}_c\n\\
    &&\quad\quad\quad+(\gamma^\mu G^{(2)}_b+\gamma^\mu G^{(2)}_c
    +H^{\mu,(2)}_b)e^{-i(\al{2}+\al{3}) p_1\cdot
    \ti{q}}\biggr\}\n\\
    &&\quad\Lambda^2_{eff}\biggl\{\Bigl(G^{(1)}_{b'} +H^{\mu,(1)}_{b'}
    \Bigr)
    e^{-i(\al{2}+\al{3}) p_1\cdot \ti{q}}+\Bigl(G^{(2)}_{b'} +H^{\mu,(2)}_{b'} \Bigr)
    e^{-i(\al{2}+\al{3}) p_1\cdot \ti{q}}
    e^{p_1\cdot p'_1}\biggr\}\Biggr]
\end{eqnarray}
The last expressions contain two types of terms, both proportional
to $\tilde{q}^2$. Since ${\Lambda^2}{\tilde{q}^2} \ll 1$, in the
IR limit, this term is totally irrelevant. One should note that
the limits taking ${\Lambda^2}\to\infty$ and $q\to
0$, is very important in our arguments. The fully renormalized
vertex function is then,
\begin{eqnarray} \l{ren}
    \Gamma^\mu_{(R)}&=&\Gamma^\mu_{1(R)}+\Gamma^\mu_{2(R)}\n\\
    &=&\frac{-\alpha\phs}{\pi}  \inta \n\\
    &&\Biggl[
    \frac{(\gamma^\mu G^{(1)}_a + H^{\mu,(1)}_a)e^{-i(\al{2}+\al{3}) p_1\cdot
    \ti{q}}}{a}+\frac{(\gamma^\mu G^{(2)}_a + H^{\mu,(2)}_a)(1-e^{i(\al{2}+\al{3}) p_1\cdot
    \ti{q}})e^{-ip_1\cdot p'_1}}{a'}\n\\
    &&\quad-2\gamma_E \biggl\{(\gamma^\mu G^{(1)}_b+\gamma^\mu G^{(1)}_c +H^{\mu,(1)}_b)
    e^{-i(\al{2}+\al{3}) p_1\cdot
    \ti{q}} +\gamma^\mu G^{(2)}_c\n\\
    &&\quad\quad\quad+(\gamma^\mu G^{(2)}_b+\gamma^\mu G^{(2)}_c
    +H^{\mu,(2)}_b)e^{-i(\al{2}+\al{3}) p_1\cdot
    \ti{q}}\biggr\}\Biggl]
\end{eqnarray}
This result is the same as that of I.F.Riad and
M.M.Sheikh-Jabbari. \ct{ja}.

The recent paper \cite{Brown:2001mg,kino} shows the 1.6 deviation
of the theoretical muon anomalous magnetic moment in the Standard Model (SM)
from the experimental data, $a^{\rm exp}_{\mu}-a^{\rm
SM}_{\mu}=426(165)\times 10^{-11}$. This result has been treated
as an indication of new physics and caused extensive interest in
many articles. We study the noncommutative QED up to 1-loop level
and correction on muon anomalous magnetic moment due to
noncommutativity.
The noncommutative QED contribution to $ a_\mu$ follows.
\begin{equation} a^{exp}_\mu - \tilde{a}{^{NC}_\mu}=a^{exp}_\mu -
(a^{QED}_\mu + \delta \tilde{a}{^{NC}_\mu})
\end{equation}
The $\tilde{a}{^{NC}_\mu}$ is obtained by the noncommutative QED
($\tilde{a}{^{NC}_\mu}=a^{QED}_\mu + \delta
\tilde{a}{^{NC}_\mu}$).

From now on, we will study the
noncommutative effect to the anomalous magnetic moment.
Up to the one loop approximation,
$\Gamma^\mu_{(R)}$ can be expanded as  functions of $q^2,
p_1.\tilde{q}~~\textrm{and}~~ \gamma.\tilde{q}$ with some coefficients
\begin{equation}
    \label{final}
    \Gamma^\mu_{(R)}=A^{\prime}\gamma^\mu + B^{\prime}(p'+p)^\mu + C\tilde{q}^\mu +
    D\gamma^\mu p.\tilde{q} +E(p'+p)^\mu\gamma.\tilde{q} .
\end{equation}
The coefficients (from $A^{\prime}$ to $E$)  are functions of
$G^i_{(a,b,c)}$ and $H^i_{(a,b,c)}$ in equation (\ref{ren}).
%======================================%
%<<<<<<<<<<<<< MOMENT >>>>>>>>>>>>>>>>>%
%======================================%
%\subsection{Anomalous Dipole Moment in NCQED}
In the NCQED, the coefficients, $C$ and $D$ can give a
contribution to the magnetic moment. In the low momentum limit\ct{ja},
the contribution of $D \gamma^\mu p.\tilde{q}$ can be ignored
and only $C$ gives a main contribution to the magnetic
moment. From the effective interaction potential with the external
magnetic field $V(x) = - <\mu> \cdot B(x)$, the magnetic moment is
given by
\begin{equation}
<\mu> = \frac{c}{i} \theta ,
\end{equation}
which is exactly the same result of Ref. \ct{ja}.

The vertex function compatible with the Ward
identity can be rewritten as
\begin{equation}
  \Gamma^\mu=A\gamma^\mu + B(p'_1+p_1)^\mu + C\ti{q}^\mu , 
\end{equation}
where $A$ is a function of $A^{\prime}$ and $D$ and $B$ a function
of $B^{\prime}$ and $E$, respectively.
In the high momentum limit,
the contribution linearly proportional to
$\theta$ and the momentum $p$ as well as the previous one evaluated
in the low momentum limit, must be considered. To evaluate the
contribution of the former,
we use the form factors $F_1(q^2)~~\textrm{and}~~F_2(q^2)$, which is expressed
as functions of $A~~\textrm{and}~~B$ using the Gordon identity and
then the invariant matrix element $M$ is given by
\begin{equation}
iM = -ie \xi^\dagger \left( - \frac{-\sigma_k}{2m} [F_1(0)+F_2(0)] \right) \xi
      \tilde{B}^k (q).
\end{equation}
Note that $M$ can be considered as the Born approximation to the scattering
of the electron with the potential
\begin{eqnarray}
iM&=&-2mi\hat V({\bf q})=-2mi\Big(-\mu^k\hat
B^k\Big)\nonumber\\&=&-2miie\theta
G_1\hat B^3\\
\mu^1=\mu^2=0&,&\mu^3(\theta)=-ieG_1\theta ,
\end{eqnarray}
where only $\mu^3$ component appears due to our choice of
noncommutativity between $x^1$ and $x^2$.
From (22), the noncommutative correction to the
magnetic moment is obtained from the following
\begin{eqnarray}
\widetilde{F_2}&=&-{\frac{\alpha}{\pi}} {\phs} {\inta} \; [A + B]\n \\
 &A&=\frac{m^2 \alpha_{1}(\alpha_{2} +  \alpha_{3}) e ^{i(\alpha_2 + \alpha_3)p
 \cdot \tilde{q}}}{\alpha_1\mu^2 + (\alpha_2 + \alpha_3)^2 m^2 - \alpha_2\alpha_3
 q^2}\n \\
 &B&=\frac{m^2 \alpha_1(\alpha_2 + \alpha_3)(1-e ^{(\alpha_2 +
 \alpha_3)p\cdot\tilde{q}}) e^{-i(p_1\cdot\tilde{p_1}+p_2\cdot\tilde{p_2})}}{m^2(\alpha_1
 - \alpha_2 - \alpha_3) + \mu ^2 (\alpha_2 + \alpha_3 ) + m^2
 (\alpha_2 + \alpha_3) + m^2 (\_2 + \alpha_3)^2
 -\alpha_2\alpha_3 q^2} ,
 \end{eqnarray}
where $A$ is the contribution of the first diagram and $B$ denotes the effects of
the second
diagram (Figure \ref{vtf}.). Using the following approximation $e^{iab} \approx 1+
iab + (\frac{iab}{2})^2 + \cdots $, the above equation can be rewritten as
\begin{eqnarray}
\widetilde{F_2}&=&-{\frac{\alpha}{\pi}} {\phs} {\inta} \;
     [\widetilde{A} + \widetilde{B}]\n \\
& \widetilde{A}&=
     \frac{m^2 \alpha_1 (\alpha_2 + \alpha_3 )[1-\frac{1}{2}(\alpha_2 +
     \alpha_3)^2 (p\theta q)^2]}{\alpha_1 \mu^2 + (\alpha_2 + \alpha_3 )^2
     m^2 - \alpha_2 \alpha_3 q^2 }\n\\
&\widetilde{B}&= - \frac{m^2 \alpha_1(\alpha_2 + \alpha_3)
     \left[ (\alpha_2+\alpha_3) (p\theta q) + \frac{1}{2}
     (\alpha_2 + \alpha_3)^2 (p \theta q)^2 \right]}
     {m^2(\alpha_1
     - \alpha_2 - \alpha_3) + \mu ^2 (\alpha_2 + \alpha_3 ) + m^2
     (\alpha_2 + \alpha_3) + m^2 (\alpha_2 + \alpha_3)^2
     -\alpha_2\alpha_3 q^2}
\end{eqnarray}
where we ignore the imaginary parts and higher order terms of $\theta$.
Before starting the calculation for $\widetilde{F_2}$, the
$\theta$ independent magnetic moment which comes from the ordinary
QED, can be derived from the $\theta$ independent terms in $\widetilde{A}$ and
here we omit the evaluation of the magnetic moment of QED. Since
we want to consider the correction of the magnetic moment caused
by the noncommutativity and related to the first order of
$\theta$, from now on we will pay attention to the term which linearly depends on
$\theta$ in $\widetilde{B}$. Since the photon mass, $\mu$ was
introduced as a IR cutoff for removing the divergence
due to the zero momentum of the photon,
we will set $\mu$ very small ($\mu \rightarrow 0$).
In the case of the soft photon ($q \rightarrow 0$),
a leading noncommutative correction term $\widetilde{F_2} (\theta \ne 0)$
reads
\begin{eqnarray}
\widetilde{F_2} (\theta \ne 0) &=& \frac{\alpha ( p \theta q )}{\pi}
     {\int_0}^1 d\alpha_2 d\alpha_3 \;
      \frac{ \{1- (\alpha_2 + \alpha_3 )\}
      (\alpha_2 + \alpha_3 )^2 }{ (\alpha_2 + \alpha_3 )^2
       - (\alpha_2 + \alpha_3 ) + 1 } \n\\
        &=& -0.074 \times \frac{\alpha ( p \theta q )}{\pi} .
\end{eqnarray}
This noncommutative correction term linearly depends on the photon momentum
$q$ and the noncommutative parameter $\theta$ and is important in
the higher momentum limit. Therefore the total magnetic moment is
summarized as the following
\begin{equation}
\langle\vec\mu\rangle_{tot} = \langle\vec\mu\rangle_0
              + \vec{\mu}_{corr} (\theta) ,
\end{equation}
where $\langle\vec\mu\rangle_0$ is the magnetic moment coming from
the ordinary QED. The noncommutative correction term $\vec{\mu}_{corr} (\theta)$
is given by
\begin{equation}
\vec{\mu}_{corr} (\theta) = \frac{\alpha
           \gamma_{Euller}}{6\pi} e m \vec{\theta}
         -0.074 \frac{e\alpha (p\theta q)}{\pi m}
         \xi^\dagger{\vec\sigma\over2}\xi .
\end{equation}
Here the first term is the leading noncommutative correction, which is consistent with
the result in Ref. \ct{ja}. The second one is
derived in the high momentum limit and so its effect in the low momentum limit can be
ignored.
In Ref. \ct{wang}, it was argued that these kinds of
noncommutative corrections can make
the SM prediction of the anomalous magnetic moment close to the experimental data.

%======================================%
%<<<<<<<<< g-2>>>>>>>>>>>>>%
%======================================%
%\section{Noncommutativity and the anomalous Magnetic Moment of the Muon}

%======================================%
%<<<<<<<<< IR DIVERGENCES >>>>>>>>>>>>>%
%======================================%
\section{Interpretation of IR divergences for the one loop level}
        \label{irdiv}
    \setcounter{equation}{0}

In the previou section we evaluated the NCQED process up to one
loop diagrams. We confirmed that the vacuum polarization diagrams
have no IR divergences while the soft bremsstrahlung diagrams have
the similar IR properties as in ordinary QED. In equation
(\ref{final}), Gordon decomposition is modified with extra piece
in NCQED. By renormalization condition for one-loop correction in
NCQED, the modified form factor $F^{'(ren)}_1$ \footnote{F' is
denote noncommutative effect(')} is ,
\begin{eqnarray}
    F^{'(ren)}_1(q^2)=F'_1(q^2)-F'_1(0)
\end{eqnarray}
Now let us confront with the IR divergence in our result
(\ref{ren}) for $F^{'(ren)}_1(q^2)$ in the vertex function. The
calculation for $F^{'(ren)}_1(q^2)$ is much more difficult.
However, it will be important in resolving the question of the IR
divergence, which we found in the discussion of bremsstrahlung.
We will find that the IR divergence coming from the
bremsstrahlung diagram and $F^{'(ren)}_1(q^2)$ cancel exactly
even for finite noncommutativity $\theta$. Although the
calculation of $F^{'(ren)}_1(q^2)$ is difficult, one can extract
useful information from by taking the limit as $\mu$ becomes
small. Then $F^{'(ren)}_1(q^2)$ integration in (\ref{final})
splits up into some pieces:
\begin{equation}
    F^{'(ren)}_1(q^2)=\sum^N_{i=1}P_i + \cdots
\end{equation}
where the ellipsis represents constant terms.

 In the IR limit,
$\mu\rightarrow 0$, the dominant part, is
\begin{eqnarray}
  F^{'(ren)}_1(q^2)&=&\frac{-\alpha\phs}{\pi}  \inta \n \\
    &&\quad\times\frac{( p_1\cdot{p_1}^{'})e^{-i(\al{2}+\al{3}) p_1\cdot
     \ti{q}}}{-\alpha_{2}\alpha_{3}q^{2}+(\alpha_{2}+\alpha_{3})^{2}m^{2}+\alpha_{1}\mu^{2}}
\end{eqnarray}
The result for the ordinary QED is in the same form except the
tildes.
%form.
%<<<<<<<<<<<<<<<<<<<<<<<<<REVISED>>>>>>>>>>>>>>>>>>>>>>>>>>>>>>>%
%\textit{This part is applied to the ordinary case except tilde
%form. }
%<<<<<<<<<<<<<<<<<<<<<<<<<REVISED>>>>>>>>>>>>>>>>>>>>>>>>>>>>>>>%

Under the limit $e^{i{x\over2}p_1 \tilde q}\approx
1+i{x\over2}p_1 \tilde q+\cdots $ we find $F'_1$
\begin{eqnarray}
 F^{'(ren)}_1(q^2)&=&{-\alpha\over\pi}\int_0^1
    d\alpha_3\int_0^1
    dx{x(m^2-{q^2\over2})\over{x^2\{(m^2-(\alpha_3-\alpha_3^2)q^2\}+\mu^2(1-x)}}\nonumber\\
    &=&{-\alpha\over2\pi}\left(m^2-{q^2\over2}\right)\bigg[\int_0^1
    d\alpha_3{\log\{m^2-(\alpha_3-\alpha_3^2)q^2\}\over{m^2-(\alpha_3-\alpha_3^2)q^2}}\nonumber\\
    &&- \log{\mu^2\over
    m^2}\int_0^1
    d\alpha_3{1\over{m^2-(\alpha_3-\alpha_3^2)q^2}}\bigg]\n \\
    &\approx&
    \frac{\alpha}{\pi}\Biggl[\biggl(\log\frac{\mu}{m}+1\biggr)(\theta
    \coth\theta -1)-2\coth \int_0^{\theta\over 2}d\phi \phi
    \tanh\phi -\frac{\theta}{4}\tanh\frac{\theta}{2}\Biggr]
\end{eqnarray}
where, \begin{eqnarray}
    p'_1\cdot p_1=m^2\cosh\theta; \; q^2=-4m^2\sinh^2(\theta/2)
\end{eqnarray} For $|q^2|\gg m^2$ we find, \begin{eqnarray}
    \gamma^\mu + \Gamma^\mu \sim
    \gamma^\mu\biggl\{1-\frac{\alpha}{\pi}\log\frac{m}{\mu}\Bigl[\log\Bigl(\frac{-q^2}{m^2}\Bigr)
    -1+O(\frac{m^2}{q^2})\Bigr]\biggr\}
    \end{eqnarray}
Plugging all this into cross-section formula, we now find our
final result,
\begin{equation}
    \frac{d\sigma}{d\Omega}(p\rightarrow
    p')=\frac{d\sigma}{d\Omega}_0
    \Biggl[1-\frac{\alpha}{\pi}\log\Bigl(\frac{-q^2}{m^2}\Bigr)
    \log\Bigl(\frac{-q^2}{\mu^2}\Bigr)+O(\alpha^2)\Biggr]
\end{equation}
%Comparing the two amplitudes,
We recall that bremsstrahlung amplitude in Eq.({\ref{so}}) in
limit $|q^2|\gg m^2$
\begin{eqnarray}
    d\sigma(p\rightarrow
    p'+\gamma)
    &\approx&d\sigma(p\rightarrow p') \cdot\frac{\alpha}{\pi}
    \log\left(\frac{-q^2}{\mu^2}\right)
    \log\left(\frac{-q^2}{m^2}\right)
\end{eqnarray} In fact, neither the elastic cross section nor the soft
bremsstrahlung cross section can be measured individually; only
their sum is physically observable. In any experiment, a photon
vdetector can detect photons only down to some minimum limiting
energy $E_l$. The probability that a scattering event occurs and
this detector does not see a photon is the sum.
\begin{equation}
    \frac{d\sigma}{d\Omega}(p\rightarrow p')+
    \frac{d\sigma}{d\Omega}(p\rightarrow p'+\gamma(k<E_l)) \equiv
    \left(\frac{d\sigma}{d\Omega}\right)_{\scriptstyle{measured}}
\end{equation}
Clearly, we find a finite,convergent result independent of
$\mu^2$, as claimed.
\begin{eqnarray}
  \frac{d\sigma}{d\Omega}(p\rightarrow
    p')_{\scriptstyle{measured}}\approx \frac{d\sigma}{d\Omega}_0
    \Biggl[1-\frac{\alpha}{\pi}\log\Bigl(\frac{-q^2}{m^2}\Bigr)\log
    \Bigl(\frac{-q^2}{E^2_{l}}\Bigr)+O(\alpha^2)\Biggr]
\end{eqnarray}

%======================================%
%<<<<<<<<<<< DISCUSSION  >>>>>>>>>>>>>>%
%======================================%
\section{Discussion}
        \label{dis}
    \setcounter{equation}{0}

In this work we have analysed some aspects of NCQED up to one
loop level. The diagrams for NCQED contain non-abelian type
diagrams. There are additional non-abelian type diagrams in the
photon vacuum polarization, and electron-photon interaction
vertex. All the UV divergences can be subtracted away by the same
local counterterms as in the ordinary QED. The main analysis of
this work is for the IR divergence.

We  analysed the soft bremsstrahlung diagrams, which is
correlated to the IR divergence of the vertex function.

First of all, the IR divergence of the soft bremsstrahlung
diagrams in the NCQED at finite noncommutativity are the same result
as that in the ordinary QED. The IR divergences of the
bremsstrahlung is shown to be cancelled out by divergence of  vertex
function in NCQED also. In vacuum polarization diagrams, there is
no IR divergence as in QED.

In section \ref{verfun}, we performed explicit calculation of the
vertex function for the photon-electron at one loop level. In
that case, it contribute to the anomalous magnetic moment and
there is a generic feature of noncommutative field theory, UV/IR
mixing.

In vertex function, we argued that the photon itself, similar to
the moving noncommutative electron,
 shows some electric dipole effect and
 magnetic dipole moment of electron has now two
parts; one is spin dependent part which will not receive any further
corrections due to the noncommutativity and the other is spin
independent, being proportional to $\theta$. In this paper, we have
calculated all noncommutative corrections proportional to $\theta$.

We have found cancellation of the IR divergence of the electron
vertex function  by the soft bremsstrahlung in the ordinary QED.
In the NCQED case, Feynman diagrams show additional non-Abelian
typed diagram from the vertex function and vacuum polarization.
Nevertheless IR divergences of all diagrams the same results as
that of the ordinary QED.

\vspace{.5cm} \noindent {\large\bf Acknowledgments} 
This work was supported by the Korea Research Foundation, Grant
No. KRF-2001-DP0083. BHL is also supported by the Sogang University
Research Grant in 2001. 
%The work is
%supported in part by the Basic Research Program of the Korea
%Science and Engineering Foundation Grant No. 1999-2-112-001-5 and
%by BK21 Project No. D-0055.
\\[.2cm]

%\vs{50mm}
%%%%%%%%%% Macro for References %%%%%%%%%%%%%%%%%%%%%%%%%%%%%%%%
\newcommand{\J}[4]{{\sl #1} {\bf #2} (#3) #4}
\newcommand{\andJ}[3]{{\bf #1} (#2) #3}
\newcommand{\AP}{Ann.\ Phys.\ (N.Y.)}
\newcommand{\MPL}{Mod.\ Phys.\ Lett.}
\newcommand{\NP}{Nucl.\ Phys.}
\newcommand{\PL}{Phys.\ Lett.}
\newcommand{\PR}{Phys.\ Rev.}
\newcommand{\PRL}{Phys.\ Rev.\ Lett.}
\newcommand{\CMP}{Comm.\ Math.\ Phys.}
\newcommand{\JMP}{J.\ Math.\ Phys.}
\newcommand{\JHEP}{J.\ High \ Energy \ Phys.}
\newcommand{\PTP}{Prog.\ Theor.\ Phys.}
\newcommand{\ib}{{\it ibid.}}
\newcommand{\hep}[1]{{\tt hep-th/{#1}}}
%%%%%%%%%%%%%%%%%%%%%%%%%%%%%%%%%%%%%%%%%%%%%%%%%%%%%%%%%%%%%%%%%%%
%%%%%%%%% References %%%%%%%%%%%%%%%%%%%%%%%%%%%%%%%%%%%%%%%%%%%%%%
%======================================%
%<<<<<<<<<<<<< Reference >>>>>>>>>>>>>>%
%======================================%

\end{document}